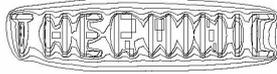

# Investigation of Micro Porosity Sintered wick in Vapor Chamber for Fan Less Design

C. S. Yu*, W. C. Wei, S. W. Kang
Department of Mechanical and Electro-Mechanical Engineering, Tamkang University
151 Ying-Chuan Rd., Tamsui, 25137, Taiwan, R.O.C.
Tel: 886-2-26215656 Ext.2613; Fax: 886-2-26209745

**Abstract**

Micro Porosity Sintered wick is made from metal injection molding processes, which provides a wick density with micro scale. It can keep more than 53 % working fluid inside the wick structure, and presents good pumping ability on working fluid transmission by fine infiltrated effect. Capillary pumping ability is the important factor in heat pipe design, and those general applications on wick structure are manufactured with groove type or screen type. Gravity affects capillary of these two types more than a sintered wick structure does, and mass heat transfer through vaporized working fluid determines the thermal performance of a vapor chamber. First of all, high density of porous wick supports high transmission ability of working fluid. The wick porosity is sintered in micro scale, which limits the bubble size while working fluid vaporizing on vapor section. Maximum heat transfer capacity increases dramatically as thermal resistance of wick decreases. This study on permeability design of wick structure is $0.5 - 0.7$, especially permeability (R) = 0.5 can have the best performance, and its heat conductivity is 20 times to a heat pipe with diameter ($\Phi$) = 10mm. Test data of this vapor chamber shows thermal performance increases over 33 %.

*Key word*：*Vapor chamber, Sintered wick structure, Heat transfer, fan less*

## 1. INTRODUCTION

Vapor chamber is full of development and technical applications, from the developments of all the countries. This study is major in sintered wick structure which present excellent capillary on work fluid transmission. The construct supports to prevent chamber deformed after vacuum processing, and also keep a smooth plan. Based on variation of the mass of working fluid, methodology of experiment for verifying the thermal performance, test data shows the best collection of sintered wick structure using in a vapor chamber.

The relative pattern in Taiwan, 1997, wick structure combined fin that located above heat source, and working fluid recycled by gravity from cooling fin structure of vapor chamber[1]. As figure 1. The relative pattern in USA, 1998, made by Thermacore Inc. [2], vapor chamber, wick structure applied on the bottom of heat pipe, and this design added aluminum fin to increase surface of heat dissipation. There existed a problem with affection of gravity of above two patterns, which caused the thermal performance become worse when heat pipe applied on different locations. Even coefficient of thermal expansion (CTE) will result in the damage on the interface of fin and heat pipe. In widely use, nowadays heat pipe in sales market is always constrained by its dimension and contact surface to heat source. The thermal expansion problem also eliminates the application on heat pipe design [3].

This study investigates the sintered wick structure on vapor chamber, as figure 2, and working fluid return to heat source without gravity affection. Except there is no any hot spot the flat surface, the first advantage of this design is the supporting wick structure, which keep the flatness of heat pipe, and also enlarge the contact surface for heat dissipation. Limit of capillary and boiling in heat pipe blocked the cooling flow back to heat source when the wick is discontinuous in the heat pipe. A vapor chamber with sintered wick structure can tell the best performance and solve all high temperature problems. Wick structure with sintered powder presents: (Sintered powder)

a. Anti-gravity to any assembly location.
b. Produce a complicate wick structure to any design.

## 2. MANUFACTURE PROCESSES

Vapor chamber of this study is using the powder with 50 % porosity, which including a copper container and DI water as working fluid. The main processes are as follows:

1. An enclosure and clean container.

2. Procedures of vacuum and filling.

3. Mass calculation of working fluid

4. Procedures of envelope.

In this study, an enclosure container is manufactured by TIG method, and there existed residual gas inside the chamber after welding process. The sintered wick structure also left metal powder on the surface of the wick. All these residues pollute a clean chamber, and affect the enveloped performance of a vapor chamber. The Residual gas blocked the path when working fluid vaporizing, also keep the saturated vapor in condense section. Those fatal error must be taken off before all test are ready, the un-sintered metal
powder is as a barrier on the wick surface, and working fluid recycling be interrupted by this situation.

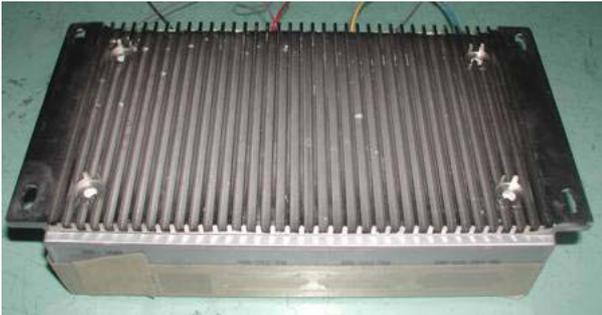

Figure1. Vapor chamber with optimal fin design

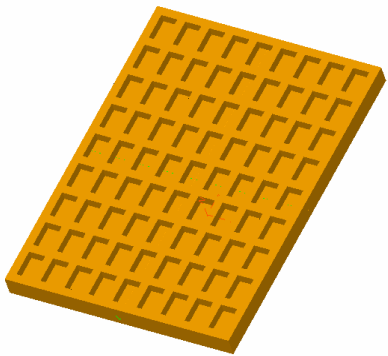

Figure2. Sintered wick structure

The processes description of vacuuming and enveloping are as follows:

1. Setup the vacuum equipment with ALCATEL vacuum pump, vacuum pipe (DN25KF), and high vacuum grease.

2. Turn the condense section down before filling the working fluid at 6.0torr, and working fluid fill into vapor chamber by pressure drop when vacuum valve opening. (one advantage of sintered wick is the vaporized working fluid will be sucked into the wick structure as pressure drop variations)

3. Second stage of vacuuming the vapor chamber, starting the recycle of cooling fluid at condense section. It prevents working fluid (De-ion water) vaporized by pressure drop variations while vacuuming the vapor chamber to $5.0 \times 10^{-1}$ torr. De-ion water has high surface tension and heat capacity which is using as the working fluid at this research, and its property is steady.

**2.1 Mass Calculation of working fluid**

In order to prevent working fluid dry out or over flooding in a vapor chamber, there is the mass formula for working fluid calculation as below. The optimal filling in a vapor chamber can improve thermal efficient and eliminate friction between the interface of vapor and liquid on wick structure, in which presents as a heat spreader at exact filling [4].

$$M_{H_2O} = V_{WV} * \rho_l \quad (2.1)$$

$$V_{WV} = V_W \varepsilon \quad (2.2)$$

$$V_W = t_W (A_{Wt}) \quad (2.3)$$

## 3. PERFORMANCE EVALUATION AND TEST

A vapor chamber works as an extreme thermal conductivity ($K_{eff}$) device, and it can transmission a huge heat flux even temperature difference is small between vapor and condense zone. We can tell heat transfer efficiency high or low from thermal conductivity of a vapor chamber [5].

Fourier's Law is used to investigate the thermal conductivity and compare with test data, for understanding the heat transfer efficiency of sintered wick structure in a vapor chamber. The equation is as follows:

$$Q_{in} = K_{eff} A_{eff} \frac{\Delta T_{measured}}{L_{eff}} \quad (3.1)$$

$$K_{eff} = \frac{Q_{in}}{A_{eff}} \frac{L_{eff}}{\Delta T_{measured}} \quad (3.2)$$

Equation (3.2) $Q_{in}$ Power input, $A_{eff}$ section of vapor chamber, $\Delta T_{measured}$ average temperature difference between vapor and condense section, $L_{ef}$ distance between vapor and condense section.

From equation (3.2), Temperature measurement of vapor and condense section is the most important parameters for performance evaluated, and obtain temperature difference ($\Delta T_{measured}$) from thermal couple measuring.($\Delta T_{measured}$), also heat loss from the wall of vapor chamber to sintered wick structure caused thermal resistance.

Components power consumption as table 1

| Chipset name | TDP(W) | Tcase(max,°C) | Tjunction(°C) |
|---|---|---|---|
| Intel Pentium M | 10 | N/A | 100 |
| Intel 82855GME | 4.3 | N/A | 110 |
| ICH | 2.3 | 105 | N/A |
| DIMM | 2.5 | 95 | N/A |

By using HP-34970 data logger and thermal couple (T-type #36) to collect temperature at each point. There are 4 points from vapor section to condense section that located on the acrylic cover, and its distance is 20 mm to each point. In order to prevent vacuum leakage on connecting the thermal couple, Silicon epoxy (Omega Omegatherm 201) applied as a seal and mounting material.

Measured temperature transmits to personal computer via RS-232 from data acquisition system, and its software helps to observe and record the temperature variation in the vapor chamber. Regarding to dimension of a vapor chamber in X-direction is 35 times to Y-direction, the influence of gravity in Y-direction is ignored. In this study, Data acquisition instrument HP-34970 can keep the precision of measurement at ±0.1 °C, and the error range of data measurement in acquisition system based on the precision of the instrument. Thermal couple (T-type) is a precise sensor on temperature measurement (test criteria is 0 °C –100 °C in this study), and it is necessary to calibrate measured temperature by referring to boiling point in atmosphere.

To maintain the surface flatness and mean thickness of a vapor chamber is the most important issue while testing, because thermal performance depends on the conductivity at each interface. Also, test system setup need to add thermal isolation on heat source to prevent error of data acquisition, which is always affected by ambient temperature.

**3.1 Thermal performance test**

Thermal performance test is based on the operating reliability and conductivity validation. Experiment spreads two parts to describe the thermal performance of this prototype [6].

An enclosure container is manufactured by TIG method, and there existed residual gas inside the chamber after welding process. The sintered wick structure also left metal powder on the surface of the wick. All these residues pollute a clean chamber, and affect the enveloped performance of a vapor chamber. The Residual gas blocked the path when working fluid vaporizing, also keep the saturated vapor in condense section.

Table2. Mass of working fluid in wick

| Vacuum pressure | methyl | WaterI | Water II | Water2-I | Water2-II |
|---|---|---|---|---|---|
|  | 21 Torr | 14 Torr | 41 Torr | 24 Torr | 37 Torr |
| Input Power | 30W~120W | | | | |
| Working fluid | 6 c.c. | 5 c.c. | 5 c.c. | 3.5 c.c. | 3.5 c.c. |
| Vaporized Temperature | < 22 °C | 28.19 °C ~ 34.8 °C | | | |

**3.2 Working fluid and vacuum pressure**

The feature of this research is using a transparent acrylic cover as an observational window [7]. We can record the operating phenomena in a vapor chamber, and compare with CFD simulation results and experimental data. According to this method to modify the model design and improve the thermal performance of a vapor chamber. Consequently, The experimental data indicate the operating situations of heat transfer and those parameters are consistent with CAE simulation results. In this study, we also provide a more efficient modified model, and show its highly thermal performance as a cooling device.

Proper thermal isolated of a heat source, which prevents heat loss during heating the vaporization area of a vapor chamber.

$$\frac{Q_{real}}{Q_{in}} = \varepsilon \quad (3.3)$$

Add ε to (3.2), then

$$K_{real} = \frac{Q_{real}}{A_{eff}} \frac{L_{eff}}{\Delta T_{measured}} = \frac{\varepsilon \cdot Q_{in}}{A_{eff}} \frac{L_{eff}}{\Delta T_{measured}} \quad (3.4)$$

**4. Result and Conclusion**

In summary, this paper shows a progression of cooling approaches that thermal engineers can use to address the demands of the growing heat dissipation levels of power electronic devices. Application of heat pipe assemblies (low heat dissipation from 10 to 125W) allow increased heat sink performance within the volume available with little potential impact on the existing system design. After cooling the electronics, which generate the most heat, there is always heat from other electronics in the system, and a sealed air-air heat exchanger is the best for dealing with the residual heat inside the system [8]. The reliability and flexibility of the heat pipe has proven to be a valuable attribute that provides the system designer with increased layout possibilities and typically improves thermal performance. The fin stack height, width, length, and fin spacing can be customized to fit the application. The benefit of temperature reduction by using micro sintered wick structure as vapor chamber prior element shows on table3.

1. Sintered wick structure with Micro particle provides excellent permeability.

2. It can keep more than 53% working fluid inside the wick structure, and a good infiltrated effect means the good pumping ability on working fluid transmission.

3. Mechanical performance verifying. (Keep surface flatness)

4. Advantages of Micro particle sintered wick structure is also lower the vaporized point of working fluid. (From 34.8°C to 28.2°C)

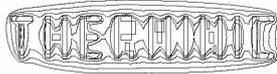
*Budapest, Hungary, 17-19 September 2007*

5. A high performance vapor chamber combine with fin, which can be a passive solution in compact system, and it can support 30W power consumption under natural convection. As figure 3.

| Chipset name | TDP(W) | Experiment result(°C) Temperature difference by using vapor chamber to copper heat sink |
|---|---|---|
| Intel Pentium M | 10 | -6 |
| Intel 82855GME | 4.3 | -5 |
| ICH | 2.3 | -3 |
| DIMM | 2.5 | -2 |

Table3. Comparison of using vapor chamber to copper plate

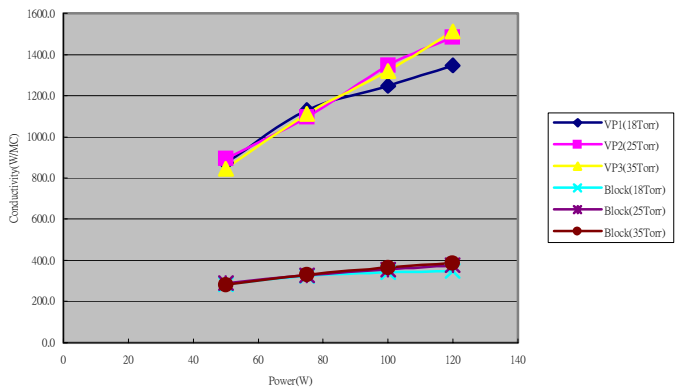

Figure 3. Thermal conductivity comparison with copper plate

**Reference**

[1] www.advantech.com.tw
[2] www.thermacore.com
[3] CFD Molding of a Therma-Base Heat Sink, Thermacore, Inc., 1998
[4] Analysis of Flow and Heat Transfer Characteristics of an Asymmetrical Flat Heat Pipe, International Journal of Heat and Mass Transfer, v35, n9, 2087, 1992
[5] Liquid Flow and Vapor Formation Phenomena in a Flat Heat Pipe, Journal of Heat Transfer Engineering, v15, n4, 33, 1994
[6] Experimental Performance of a Heat Pipe, International Communication of Heat and Mass Transfer, v26, n5, 679, 1999
[7] Vapor Pressure Distribution of a Flat Plat Heat Pipe International Communications in Heat and Mass Transfer 1996 v23 Optimal Heat Pipe Design, International Communication in Heat and Mass Transfer, v24, n3, 371, 1997
[8] D. Khrustalev, A. Faghri," Thermal characteristics of conventional and flat miniature axially grooved heat pipes", J. Heat Transfer 117, pp.1048-1054 2005